# Hidden gauge structure and derivation of microcanonical ensemble theory of bosons from quantum principles

Sumiyoshi Abe[1] and Tsunehiro Kobayashi[2]

[1]*Institute of Physics, University of Tsukuba, Ibaraki 305-8571, Japan*
[2]*Department of General Education for the Hearing Impaired,*
*Tsukuba College of Technology, Ibaraki 305-0005, Japan*

Microcanonical ensemble theory of bosons is derived from quantum mechanics by making use of a hidden gauge structure. The relative phase interaction associated with this gauge structure, described by the Pegg-Barnett formalism, is shown to lead to perfect decoherence in the thermodynamic limit and the principle of equal *a priori* probability, simultaneously.

Quantum theoretical justification of statistical mechanics is a problem of fundamental interest [1,2]. There seem to remain several important questions yet to be answered. First of all, (i) perfect decoherence has to be realized at the level of microcanonical ensemble theory, in which there is no notion of the heat bath. (ii) The principle of equal *a priori* probability should be established based on quantum dynamics. In addition, (iii) what is the quantum analog of classical ergodicity that may allow statistical description of a system? Consider the classical ideal gas, for example. The particles of the gas actually interact with each other in a complex manner. This interaction is responsible for validitiy of ergodicity but is ignored if once shifted to statistical description. Then, (iv) what is the physical property of such an interaction in quantum theory?

In this paper, we study quantum mechanical derivation of the microcanonical ensemble theory of bosons by answering to all the above questions (i)-(iv) *simultaneously*. Our discussion is based crucially on the hidden gauge structure in the system. This structure is made manifest within the framework of the Pegg-Barnett unitary-phase-operator formalism. We heuristically construct the interaction Hamiltonian of a particular form, in which all relevant complex interactions needed for statistical description are assumed to be effectively summarized. Then, we show that the eigenstate of the total Hamiltonian satisfies the principle of equal *a priori* probability and, at the same time, perfect decoherence is realized in the thermodynamic limit with the help of the hidden gauge structure. Once the microcanonical ensemble theory is obtained, derivation of the canonical ensemble theory is straightforward.

Let us consider a system of *N* identical bosons. The Hamiltonian reads

$$H_0 = \varepsilon \sum_{i=1}^{N} a_i^\dagger a_i, \tag{1}$$

where $a_i^\dagger$ and $a_i$ are the ordinary creation and annihilation operators of the *i*th boson satisfying the algebra, $[a_i, a_j^\dagger] = \delta_{ij}$, $[a_i, a_j] = [a_i^\dagger, a_j^\dagger] = 0$, and $\varepsilon$ is the energy of a single boson. According to the spirit of statistical mechanics, there are weak interactions between the bosons, which are responsible for the validity of statistical approach, but they can be ignored once the system is statistically described. We wish to find the interaction Hamiltonian, $H_I$, which represents such interactions in a simple and effective way and leads to the principle of equal *a priori* probability. For this purpose, we consider the phase degree of freedom within the framework of the Pegg-Barnett formalism [3]. We shall consider only the stationary states of the isolated system, and the problem of relaxation to equilibrium will not be discussed.

A generalization of the following discussion to non-identical particles is straightforward but brings extra complications.

The Pegg-Barnett phase state of the *i*th boson is given by

$$\left|\theta_{m_i}\right\rangle_i = \frac{1}{\sqrt{s+1}} \sum_{n_i=0}^{s} \exp(i n_i \theta_{m_i}) \left|n_i\right\rangle_i, \qquad (2)$$

$$\theta_{m_i} = \frac{2\pi m_i}{s+1} \qquad (m_i = 0, 1, 2, \cdots, s), \qquad (3)$$

where $\left\{\left|n_i\right\rangle_i\right\}_{n_i=0,1,2,\cdots,s}$ is the truncated Fock basis satisfying $a_i^\dagger a_i \left|n_i\right\rangle_i = n_i \left|n_i\right\rangle_i$ and $_i\left\langle n_i | n_i'\right\rangle_i = \delta_{n_i, n_i'}$. $\left\{\left|\theta_{m_i}\right\rangle_i\right\}_{m_i=0,1,2,\cdots,s}$ forms an orthonormal complete system in the $(s+1)$-dimensional space. $s$ is supposed to be large, but the limit $s \to \infty$ has to be taken after all quantum mechanical calculations. The unitary phase operator is given by

$$\exp(i\phi_i) = \sum_{m_i=0}^{s} \exp(i\theta_{m_i}) \left|\theta_{m_i}\right\rangle_i {}_i\!\left\langle\theta_{m_i}\right|. \qquad (4)$$

It is worth recalling that the phase operator is an anomalous object and quantum-classical correspondence between the action and angle variables is violated, in general [4].

Applying the operator in Eq. (4) on $\left|n_i\right\rangle_i$, we have

$$\exp(i\phi_i)\left|n_i\right\rangle_i = \left|n_i - 1\right\rangle_i \quad (n_i \neq 0), \qquad \exp(i\phi_i)\left|0\right\rangle_i = \left|s\right\rangle_i \quad (n_i = 0). \qquad (5)$$

It is of importance to notice that the relations in Eq. (5) remain form invariant under the following transformation:

$$\left|n_i\right\rangle_i \to \left|n_i, \theta_{m_i}\right\rangle_i = \left|n_i\right\rangle_i \exp(i\alpha_{n_i}), \qquad (6)$$

$$\phi_i \to \phi_i - \partial \alpha_{n_i} \equiv \phi_i - \left(\alpha_{n_i+1} - \alpha_{n_i}\right) = \phi_i - \theta_{m_i}, \tag{7}$$

where $\alpha_{n_i} = n_i \theta_{m_i}$. This can be seen as a kind of gauge transformation, and the role of a gauge field is played by $\phi_i$.

Taking into account this hidden gauge structure, we here present the following heuristically-constructed interaction Hamiltonian:

$$H_\text{I} = g\left[ V^\dagger V + N \sum_{i=1}^{N} |0\rangle_{i\ i}\langle 0| \right], \tag{8}$$

$$V = \sum_{i=1}^{N} \left\{ \exp\left[i\left(\phi_i - \theta_{m_i}\right)\right] - |s\rangle_{i\ i}\langle 0| \right\}, \tag{9}$$

where $g$ is a coupling constant. In Eq. (9), we have used the notational abbreviation $\sum_{i=1}^{N} A_i = A_1 \otimes I_2 \otimes \cdots \otimes I_N + I_1 \otimes A_2 \otimes I_3 \otimes \cdots \otimes I_N + \cdots + I_1 \otimes \cdots \otimes I_{N-1} \otimes A_N$ with $I_i$ being the unit operator in the space of the $i$th boson. The subtraction term in $V$ is a feature of the Pegg-Barnett formalism. $H_\text{I}$ can be thought of as the effective Hamiltonian, in which all relevant complex interactions needed for validating statistical description are summarized. Note that it essentially represents the relative phase interaction [5].

Since our purpose is to derive statistical mechanics of the system with $H_0$ and without $H_I$ containing anomalous objects, we here impose the condition that $H_I$ should vanish in the thermodynamic limit, $N \to \infty$. It turns out that this condition is fulfilled if the coupling constant is assumed to decay faster than $1/N^2$. Thus, we put

$$g = \frac{g_0}{N^{2+\delta}} \qquad (\delta > 0), \tag{10}$$

where $g_0$ is a constant independent of $N$.

For a finite value of $N$, the normalized eigenstate of the total Hamiltonian, $H = H_0 + H_I$, is found to be given by

$$\left| M; N, [\theta_m] \right\rangle = \frac{1}{\sqrt{W(M, N)}} \sum_{P\{n\}} \left| M; [n], [\theta_m] \right\rangle, \tag{11}$$

with

$$\left| M; [n], [\theta_m] \right\rangle = \bigotimes_{i=1}^{N} \left| n_i \right\rangle_i \times \delta_{n_1+n_2+\cdots+n_N, M} \exp\left( i \sum_{i=1}^{N} n_i \theta_{m_i} \right), \tag{12}$$

$$W(M, N) = \frac{(M+N-1)!}{M!(N-1)!}, \tag{13}$$

whose energy eigenvalue is given by

$$E = M\varepsilon + \frac{g_0}{N^{\delta}} \qquad (M = 0, 1, 2, \cdots). \tag{14}$$

The symbol, P, stands for permutation and therefore the summation in Eq. (11) is understood to be taken over all possible combination of $\{n\} \equiv (n_1, n_2, \cdots, n_N)$. For the sake of nontriviality, the dimensionality, $s$, should be taken to be larger than $M$.

The space of the quantum states is enlarged by the introduction of the phase variables. The state in Eq. (11) residing in such a space is found to satisfy the following normalization condition:

$$\frac{1}{(s+1)^N} \text{Tr} \sum_{m_1, m_2, \cdots, m_N = 0}^{s} |M; N, [\theta_m]\rangle\langle M; N, [\theta_m]| = 1. \tag{15}$$

Now, let $A$ be a normal physical observable, which is *independent of the phase operators with anomaly*. Then, taking Eq. (15) into account, its quantum mechanical average, $<A>$, with respect to the state in Eq. (11) converges in the thermodynamic limit as well as the Pegg-Barnett limiting procedure, $s \to \infty$, as follows:

$$<A> \to \text{Tr}(A \rho_{\text{mc}}), \tag{16}$$

where $\rho_{\text{mc}}$ is given by

$$\rho_{\text{mc}} = \frac{1}{W(M, N)} \sum_{\text{P}\{n\}} |n_i\rangle_i\, _i\langle n_i| \times \delta_{n_1 + n_2 + \cdots + n_N, M}. \tag{17}$$

This is precisely the microcanonical density matrix of the bosons with the fixed energy, $E_0 = M\varepsilon$, in which perfect decoherence and the principle of equal *a priori* probability are realized simultaneously, as desired.

Finally, the canonical ensemble theory is derived in the standard manner. The total system is divided into the objective system ($S$) and the heat bath ($B$): $N = N_S + N_B$ ($N_S \ll N_B$), $E_0 = E_{S,M_S} + E_{B,M_B}$ with $E_{S,M_S} = M_S \varepsilon$ and $E_{B,M_B} = M_B \varepsilon$ ($M_S \ll M_B$). The total state is written as

$$\left|M; N, [\theta_m]\right\rangle = \sum_{M_S, M_B} \sqrt{\frac{W(M_S, N_S)\, W(M_B, N_B)}{W(M, N)}} \\ \times \left|M_S; N_S, [\theta_{S,m}]\right\rangle_S \otimes \left|M_B; N_B, [\theta_{B,m}]\right\rangle_B \delta_{M_S+M_B, M}, \qquad (18)$$

where the states of the objective system and the heat bath are respectively given by

$$\left|M_S; N_S, [\theta_{S,m}]\right\rangle_S = \frac{1}{\sqrt{W(M_S, N_S)}} \sum_{\mathrm{P}\{n\}_S} \left|M_S; [n]_S, [\theta_{S,m}]\right\rangle_S, \qquad (19)$$

$$\left|M_B; N_B, [\theta_{B,m}]\right\rangle_B = \frac{1}{\sqrt{W(M_B, N_B)}} \sum_{\mathrm{P}\{n\}_B} \left|M_B; [n]_B, [\theta_{B,m}]\right\rangle_B, \qquad (20)$$

in the notation analogous to Eqs. (11)-(13). The canonical density matrix of the objective system is obtained by performing the partial trace over the heat bath:

$$\rho_c = \frac{1}{(s+1)^N} \mathrm{Tr}_B \sum_{m_1, m_2, \cdots, m_N = 0}^{s} \left|M; N, [\theta_m]\right\rangle\!\left\langle M; N, [\theta_m]\right|$$

$$\cong \frac{1}{Z(\beta)} \sum_{M_S} \exp(-\beta E_{S,M_S}) \times \frac{1}{W(M_S, N_S)} \sum_{P\{n\}_S} |M_S;[n]_S\rangle_{S\ S}\langle M_S;[n]_S|, \quad (21)$$

where

$$|M_S;[n]_S\rangle_S = \bigotimes_{i=1}^{N_S} |n_i\rangle_{S,i} \times \delta_{n_1+n_2+\cdots+n_{N_S}, M_S}, \quad (22)$$

$$Z(\beta) = \sum_{M_S} \exp(-\beta E_{S,M_S}), \quad (23)$$

$$\beta = \frac{\partial W(M_B, N_B)}{\partial E_{B,M_B}} = \frac{1}{\varepsilon} \ln\left(1 + \frac{N_B}{M_B}\right), \quad (24)$$

provided that the Boltzmann constant has been set equal to unity.

In conclusion, we have presented a model of the bosons with a hidden gauge structure, which may explain, in the thermodynamic limit, realization of perfect decoherence and the principle of equal *a priori* probability in the microcanonical ensemble theory.

S. A. would like to thank Professor T. Kunihiro and Yukawa Institute of Theoretical Physics, Kyoto University, for hospitality extended to him. He also thank Dr. A. K. Rajagopal for discussions.